\documentclass[preprint,superscriptaddress,showpacs,preprintnumbers,amsmath,amssymb,aps, pre]{revtex4-1}

\usepackage[utf8]{inputenc}
\usepackage[english]{babel}
\usepackage{subfigure}
\usepackage{graphicx}
\usepackage{dcolumn}
\usepackage{bm}
\usepackage{enumitem}
\usepackage{xcolor}

\begin{document}

\title{Ascertaining when a basin is Wada: the merging method}

\author{Alvar~Daza}
\affiliation{Nonlinear Dynamics, Chaos and Complex Systems Group,
Departamento de F\'{i}sica, Universidad Rey Juan Carlos, Tulip\'{a}n
s/n, 28933 M\'{o}stoles, Madrid, Spain}

\author{Alexandre~Wagemakers}
\affiliation{Nonlinear Dynamics, Chaos and Complex Systems Group,
Departamento de F\'{i}sica, Universidad Rey Juan Carlos, Tulip\'{a}n
s/n, 28933 M\'{o}stoles, Madrid, Spain}

\author{Miguel A.F. Sanju\'{a}n}
\affiliation{Nonlinear Dynamics, Chaos and Complex Systems Group,
Departamento de F\'{i}sica, Universidad Rey Juan Carlos, Tulip\'{a}n
s/n, 28933 M\'{o}stoles, Madrid, Spain}
\affiliation{Department of Applied Informatics, Kaunas University of Technology, Studentu 50-415, Kaunas LT-51368, Lithuania}
\affiliation{Institute for
Physical Science and Technology, University of Maryland, College
Park, Maryland 20742, USA}

\date{\today}

\begin{abstract}
Trying to imagine three regions separated by a unique boundary seems a difficult task. However, this is exactly what happens in many dynamical systems showing Wada basins. Here, we present a new perspective on the Wada property: \textit{A Wada boundary is the only one that remains unaltered under the action of merging the basins}. This observation allows to develop a new method to test the Wada property, which is much faster than the previous ones. Furthermore, another major advantage of the merging method is that a detailed knowledge of the dynamical system is not required.
\end{abstract}

\maketitle

\section{\label{sec:Introduction}Introduction}

Wada basins are one of those unexpected encounters that often happen in science. The story begins when a topologist named Takeo Wada tried to answer the following question: Can three or more open regions be separated by a single boundary? Our daily experience makes us think that this is impossible. It suffices to look at a common political map to realize that the boundaries separate two different regions, except perhaps some isolated points that separate three or more regions (think about the Four Corners in the USA, for example). However, Takeo Wada devised an iterative process to make this counter-intuitive situation possible, as reported by his student Kunizo Yoneyama \cite{yoneyama_theory_1917,hocking_topology_1988}. The Wada lakes were conceived in a topological context as a way to separate three connected regions in a plane by means of a continuous boundary \cite{yoneyama_theory_1917}. From a topological point of view, Wada boundaries have intriguing properties. For example, the Polish topologist Kazimierz Kuratowski showed that in the plane, continuous Wada boundaries must be indecomposable continua \cite{kuratowski_sur_1924} (though the situation in higher dimensions is quite different).

This discovery remained as a mathematical curiosity until James Yorke and his collaborators found that the basins of attraction of some dynamical systems presented the Wada property \cite{kennedy_basins_1991,nusse_saddle-node_1995}. From the dynamical point of view, the most interesting feature of Wada basins is the fact that an arbitrarily small perturbation of a system with initial conditions lying in a Wada boundary can drive it to any of the possible attractors, which implies a special kind of unpredictability \cite{daza_basin_2016}. Therefore, in this context, Wada boundaries are usually referred to as those that separate three or more basins at a time, but the basins need not to be connected. Since the earliest references to the Wada property in dynamical systems, many authors claim that the boundaries have the Wada property for disconnected basins \cite{poon_wada_1996, epureanu_fractal_1998, sweet_topology_1999, aguirre_wada_2001, vandermeer_wada_2004, daza_wada_2017}. In this work, we adopt this latter definition: Wada boundaries are those that separate three or more basins, no matter whether the basins are connected or not. Therefore, using this definition, we believe that the methodology and the results presented in this work are valid for any number of dimensions.

Despite our primary intuition, Wada basins are a common feature appearing in many dynamical systems. Since its first report, Wada basins have been found in open Hamiltonian systems \cite{aguirre_wada_2001}, ecological models \cite{vandermeer_wada_2004}, delayed differential equations \cite{daza_wada_2017}, hydrodynamical systems \cite{toroczkai_wada_1997}, and many engineering problems \cite{aguirre_unpredictable_2002,zhang_unpredictability_2012,zhang_wada_2013}. This is possible because Wada boundaries are related to iterative processes and fractal structures, which are a common feature in the basins of nonlinear dynamical systems \cite{aguirre_fractal_2009}.

So far, two methods have been proposed to determine when the basins of a system possess the Wada property. The first one was developed by Nusse and Yorke \cite{nusse_wada_1996,nusse_fractal_2000}, and involved the computation of the unstable manifold of a saddle point of the basin boundary. This method requires a detailed knowledge of the system and the computation of unstable trajectories, which can be cumbersome in many cases. Indeed, many papers \cite{aguirre_wada_2001, aguirre_unpredictable_2002, toroczkai_wada_1997, portela_fractal_2007} are devoted just to determine whether the Nusse-Yorke condition is fulfilled in a particular dynamical system and for a certain set of parameters. Years after the original works by Yorke and collaborators, Daza et al. \cite{daza_testing_2015} developed a grid method based on the successive refinement of the grid in order to determine whether all the boundary points were Wada points (points that separate three or more basins at a time). This latter method can be automated and used for every dynamical system. As a drawback, it requires the precise computation of new trajectories at very high resolutions. Although it supposes a qualitative and quantitative improvement with respect to the Nusse-Yorke method, the grid method needs several hours or even days of parallel computation to check the Wada property in a given dynamical system. In this paper, we present a new method to determine when a basin is Wada, which is founded on the observation that: \textit{A Wada boundary is the only one that remains unaltered under the action of merging the basins}. This new method, that we call \textit{the merging method}, can test the Wada property in a few seconds, and furthermore it does not require a detailed knowledge of the system. Thus, the merging method supposes a new quantitative and qualitative leap with respect to the previous available methods to test the Wada property.

The paper is organized as follows. In Sec.~\ref{sec:Merging}, we explain how Wada boundaries can be defined as the only ones that remain unaltered after the action of merging the basins. In Sec.~\ref{sec:MergingMethod}, based on the previous definition, a new method to test the Wada property is presented. Section~\ref{sec:Analysis} is devoted to the detailed analysis of the merging method using different model examples. Finally, we discuss the results and present the main conclusions of the paper in Sec.~\ref{sec:Conclusions}.


\section{\label{sec:Merging}Merging basins}

The set of all initial conditions leading to a particular attractor is called the basin of attraction of a dynamical system. We will focus on a very special set of initial conditions called the boundary. A point $p$ is in the \emph{boundary} of a basin $B_i$ if  $\forall\varepsilon>0$, the open ball centered in $p$ of radius $\varepsilon$, $b(p,\varepsilon)$, is such that $b(p,\varepsilon)\cap B_i \neq \emptyset$ and $b(p,\varepsilon)\cap  {B_i}^\complement \neq \emptyset$, where ${B_i}^\complement$ is the complement of $B_i$. If the point satisfies the previous condition for all the basins $B_i$ with $N_a \geq 3$ basins of attraction, we call it a \emph{Wada point}. If all the boundary points are Wada points, then the basin of attraction has the \emph{Wada property}, and we call it a Wada basin.

However, we can formulate the Wada property in slightly different terms. Assume we have $N_a \geq 3$ basins of attraction. Now, we want to determine the boundary $\partial B_i$ of each basin $B_i$, but instead of using its complement ${B_i}^\complement$ to determine which points belong to the boundary, we will say that a point $p$ is in the boundary if it is arbitrarily close to $B_i$ and also arbitrarily close to at least one of the other basins $B_j$. That is, $p$ is in the boundary $\partial B_i$ if  $\forall\varepsilon>0$, the open ball centered in $p$ of radius $\varepsilon$, $b(p,\varepsilon)$, is such that $b(p,\varepsilon)\cap B_i \neq \emptyset$ and $b(p,\varepsilon)\cap  \bigcup\limits_{j\neq i} B_j \neq \emptyset$. Then, we determine each boundary $\partial B_i$ as the boundary between a basin $B_i$ and all the other merged basins $\bigcup\limits_{j\neq i} B_j$, so that we end with as many different boundaries as different possible attractors, $i=1,\ldots, N_a$. Thus, all the boundaries created following this previous procedure and the boundary of the original basins corresponding to the $N_a$ attractors are exactly the same $\partial B_i=\partial B_j$ for $\forall i \neq j, i=1,\ldots, N_a$, if and only if the system is Wada.


\begin{figure*}
\begin{center}
\includegraphics[width=\textwidth]{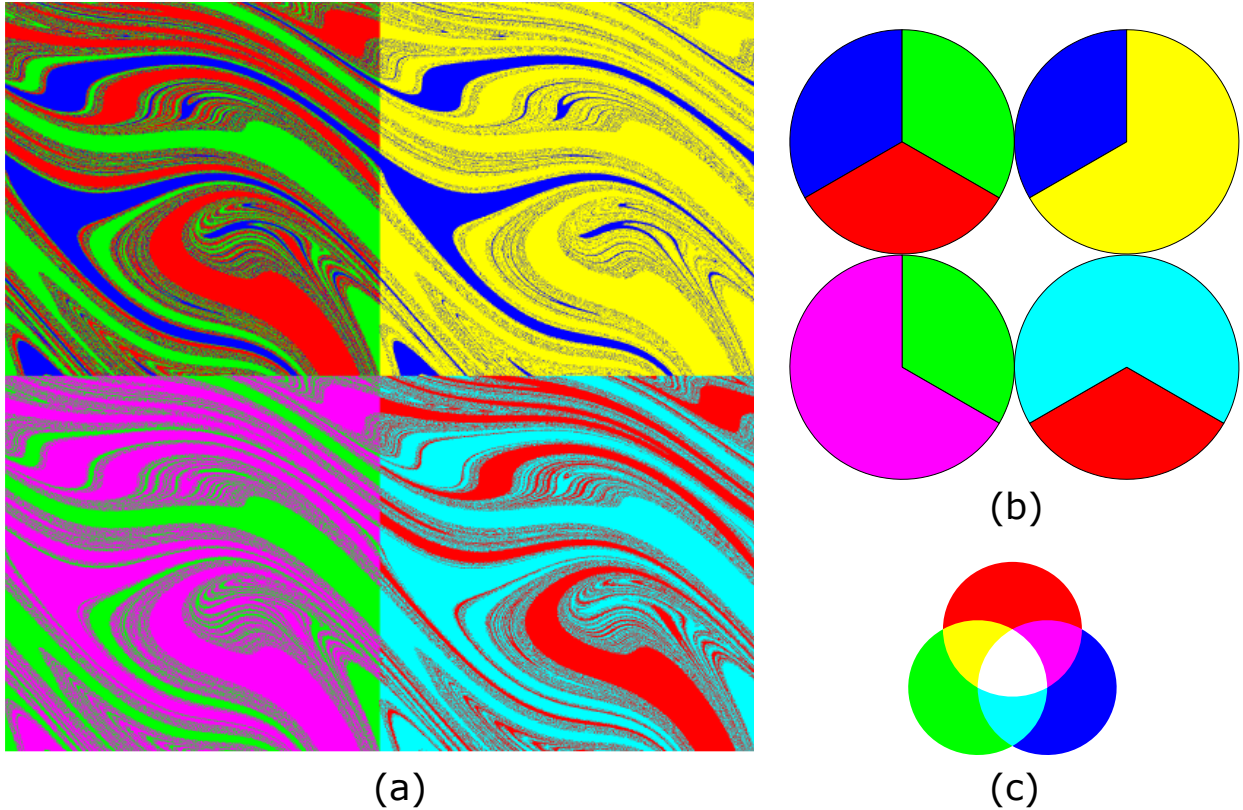}
\end{center}
\caption{\label{fig:Wadarhol} \textbf{Merging Wada basins.} The time-$2\pi$ (Poincar\'e) map of the forced damped pendulum defined by $\ddot{x}+0.2\dot{x}+\sin x=1.66\cos t$ possesses three attractors, and consequently its phase space $(x,\dot{x})$ contains three basins. This system verifies the Wada property~\cite{kennedy_basins_1991}. (a) The top-left panel (in red, green and blue) represents the Wada basins, where one color corresponds to one basin. The other three panels are the result of the action of merging the basins: we merge two colors, and keep the third unchanged. (b) In the top-left panel a disk is divided in three colors. The other three panels show the action of merging in this non-Wada picture. (c) The color-code of the merged basins can be inferred from the bottom-right picture: yellow=red+green, magenta=blue+red, cyan=blue+green.}
\end{figure*}


The two previous definitions of Wada basins are completely equivalent. However, the second definition emphasizes the striking idea that Wada basins can be merged and the boundary will still remain unaltered. To be more precise, given $N_a\geq3$ Wada basins, it is possible to merge up to $N_a-1$ without any change in the boundary (if we merge the $N_a$ basins then there would be only one basin and the boundary would be lost). This notable effect is better appreciated when Wada boundaries are compared to non-Wada boundaries. The time-$2\pi$ (Poincar\'e) map of the forced damped pendulum defined by $\ddot{x}+0.2\dot{x}+\sin x=1.66\cos t$ possesses three attractors, and consequently its phase space $(x,\dot{x})$ contains three basins. This is a paradigmatic system showing Wada basins~\cite{kennedy_basins_1991}. In the top-left panel of Fig.~\ref{fig:Wadarhol}-(a), we display the original three-colored Wada basins of the forced damped pendulum described in \cite{kennedy_basins_1991}. The other three plots show the result of merging the basins according to the color code sketched in Fig.~\ref{fig:Wadarhol}-(c) (yellow=red+green, magenta=blue+red, cyan=blue+green). It is important to notice that each color represents a different basin, being impossible to establish a one-to-one correspondence between basins of different colors. However, even though the basins are different, the boundaries are the same for the four panels of Fig.~\ref{fig:Wadarhol}-(a).


If we look at the colored disks in Fig.~\ref{fig:Wadarhol}-(b), we can see how the action of merging affects usual (non-Wada) basins. Here we can clearly notice that the boundaries change under the action of merging the basins. In fact, the center of the disk is the only point that is in the boundary of the four panels, so that it is a Wada point.

Despite the abundant research devoted to Wada basins, the effect that the Wada boundaries remain unaltered after the action of merging the basins has been unnoticed. In the next sections, we will use it to develop a new way of testing Wada basins in dynamical systems.

\section{Merging basins to test the Wada property}
\subsection{\label{sec:MergingMethod}Description of the merging method}
The property that we have just described, that is, that Wada basins can be merged without any change in their boundary, can be used to build a new method to test the Wada property. From a purely mathematical point of view, given the basins of a system, it suffices to check that the boundary remains unaltered under the merging of the basins. However, it would require an arbitrarily high resolution of the basins to guarantee that the boundaries of the merged basins are exactly the same.

\begin{figure*}
\begin{center}
\includegraphics[height=0.7\textheight]{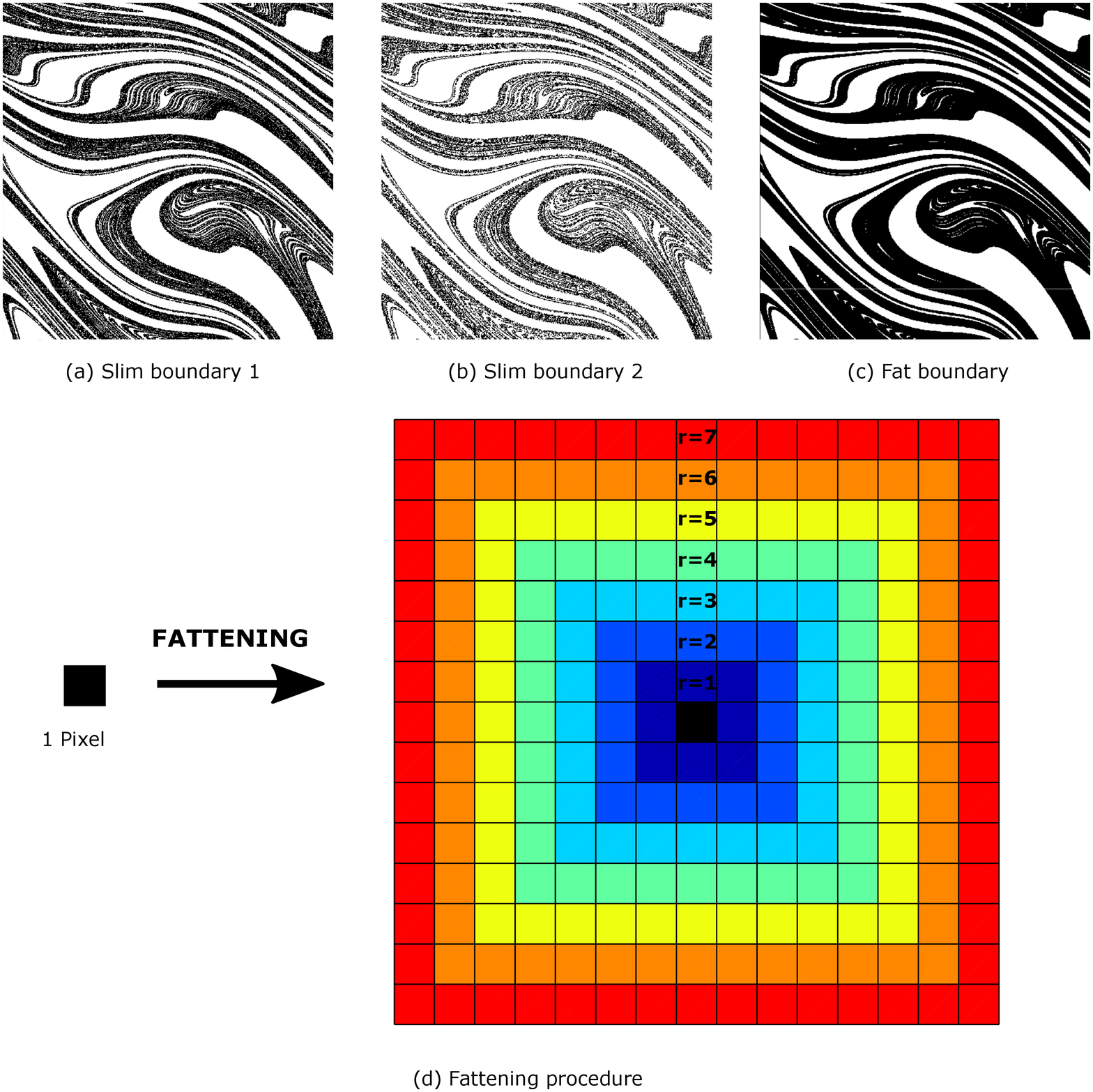}
\end{center}
\caption{\label{fig:fattening} \textbf{The fattening procedure.} (a)-(b) Even for Wada basins, the boundaries are not exactly the same for all the merged basins because of the finite resolution. (c) To avoid this effect, the boundaries must be fattened. (d) In the fattening procedure, each pixel in the boundary is substituted by a fat pixel of radius $r$. In the plot, each color corresponds to a different radius in the Chebyshev or chessboard distance. We call $r$ the \textit{fattening parameter}.}
\end{figure*}

Usually, the basins are computed by means of a regular grid of finite size. In this approach, every pixel of the grid has a linear size $\varepsilon$ and contains only one corresponding initial condition, in such a way that the fate of this initial condition determines the color of the pixel. Therefore, the computation of the boundaries is limited by the size of the pixel $\varepsilon$. In Fig.~\ref{fig:fattening}-(a)-(b), we can see that the computed boundaries of the merged basins, which we call \textit{slim boundaries}, are not exactly the same, even though they are Wada boundaries. It can be observed at naked eye that although the boundaries seem similar, they are not strictly identical. It is noticeable that the boundary depicted in Fig.~\ref{fig:fattening}-(a) is \textit{thicker} than the boundary depicted in Fig.~\ref{fig:fattening}-(b).

To overcome this issue, we can try to \textit{fatten} the boundaries for their subsequent comparison (see the fattened boundary of Fig.~\ref{fig:fattening}-(c)). In the fattening procedure, we replace each pixel belonging to the boundary by a \textit{fat pixel} defined by \textit{the fattening parameter $r$}. This fattening parameter is the radius of the fat pixel according to the Chebyshev metric or maximum distance metric. This metric preserves the square shape of the pixels and it is defined in the plane as $r=\max\left(\vert x_2-x_1\vert, \vert y_2-y_1\vert \right)$, where $(x,y)$ are the usual Cartesian coordinates. Sometimes this metric is also called the chessboard distance, since it represents the number of moves that a king would have to make to go from one position to another (see Fig.~\ref{fig:fattening}-(d)). The way the fattening is made can be changed according to different metrics, this is not crucial for the method. In the next section, we will analyze how the method that we are describing depends on the fattening parameter $r$. Now, let us move forward to the last part of the procedure.



For the moment, we have the original boundaries of the merged basins, the \textit{slim boundaries} $\partial B_i$, and their fattened versions, the \textit{fat boundaries} $\boldsymbol{\partial B_i}$. The final step of the procedure is to compare all the slim boundaries with all the fat boundaries. If all the slim boundaries fit in all the fat boundaries $\partial B_i \subset \boldsymbol{\partial B_j}$ $\forall i,j=1,\ldots,N_a$; then we will say that the basin is Wada. Otherwise, we will say that the system is not Wada, and the method will determine which points are Wada and which ones are not. This last step verifies if each pixel of the \textit{slim boundaries} $\partial B_i$ lies in the set $\boldsymbol{\partial B_j}$. To connect with our previous definition of a basin with the Wada property, the algorithm checks if the points $p_i$ in the boundaries $B_i$ are within a ball $b(p_j,r)$ of radius $r$ ($r$ is the fattening parameter) around the points $p_j$ of the boundary $B_j$.

In the case of partially Wada basins \cite{zhang_wada_2013}, where Wada and non-Wada boundaries coexist, we can characterize them by the Wada parameter $W_{N_a}$ defined in the grid method of Daza {\it et al.}\cite{daza_testing_2015}. This parameter $W_{N_a}$ provides the ratio of Wada points to boundary points (Wada and non-Wada), in such a way that $W_{N_a}=1$ means that the system has the full Wada property, whereas $W_{N_a}<1$ indicates only partially Wada basins. In the merging method, given a basin, we can compute the pixels lying in the boundary of that basin $n_b$, and we can also register the number of boundary points which are not Wada $n_{NW}$. Then, the Wada parameter for a fixed resolution can be calculated simply as $W_{N_a}=1-n_{NW}/n_b$.


Again, for a better understanding of the comparison between slim and fat boundaries, it is convenient to observe an example of non-Wada basins, such as the disks of Fig.~\ref{fig:Wadarhol}-(b). We would have to fatten the boundaries by a very large amount (comparable to the size of the disks) in order to make the fat boundaries able to contain the slim ones. We can conclude, as mentioned before, that the only Wada point is the center of the disk.

\begin{figure*}
\begin{center}
\includegraphics[height=0.75\textheight]{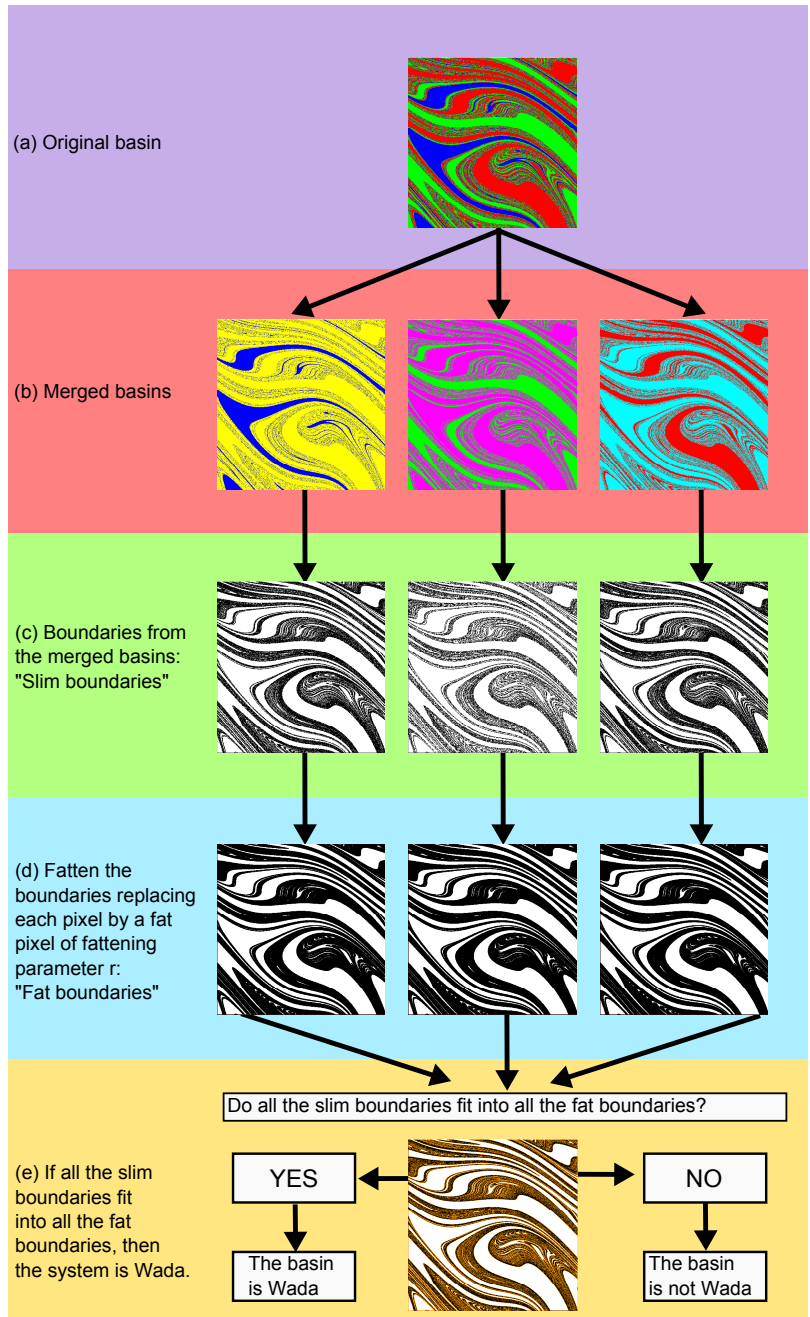}
\end{center}
\caption{\label{fig:flowchart} \textbf{Flowchart of the merging method to test Wada basins.} (a) Originally we have the picture of a basin at a given resolution. (b) We merge the basins, so that we have as many merged basins as different colors in the original basin. (c) We find the boundaries of the merged basins, which we can see they are similar but not exactly the same. (d) We fatten the boundaries using fat pixels of fattening parameter $r$. (e) We check if all the slim basins are contained in the fat basins. If this is the case, then the basin is Wada, otherwise, it is not.}
\end{figure*}

The whole procedure described before can be fully automated and the only input needed is a finite resolution basin. For basins with a resolution of $1000 \times 1000$ and three different colors (attractors), the merging method takes around two seconds to determine whether a basin is Wada running in a personal computer. This contrasts with previous methods to test Wada basins. The grid method \cite{daza_testing_2015} needs to compute new trajectories at finer resolutions, which can take several hours or even days of parallel computation in a cluster with one hundred cores. The Nusse-Yorke method \cite{nusse_wada_1996,nusse_fractal_2000} requires detailed knowledge of the dynamics of the system and, in general, it cannot be automated. In fact, many works \cite{aguirre_wada_2001, aguirre_unpredictable_2002, toroczkai_wada_1997, portela_fractal_2007} are exclusively devoted to the application of the Nusse-Yorke method to one particular system and one specific set of parameters due to the difficulty of the task. In comparison, the merging method is incredibly fast and easy, since it only requires a finite resolution basin to be applied. Furthermore, since the merging method does not need any further computation of new points, we do not even need to know the underlying dynamical system nor its parameters.

Next we summarize the steps that this merging algorithm takes, which can also be visualized in the flowchart of Fig.~\ref{fig:flowchart}.

\begin{enumerate}[label=(\alph*)]
\item At first, we have a picture of the basins at a given resolution $\varepsilon$. As we will discuss later, the higher the resolution the more reliable the determination of the Wada property will be.

\item For each basin $B_i$, we merge the other basins obtaining two-color basins of attraction made of the original basin $B_i$ and the merged basin $\bigcup\limits_{j\neq i} B_j$. By this process, we get a collection of $N_a$ pictures with only two colors.

\item We compute the slim boundaries of the merged basins $\partial B_i$. In order to do this, we can simply see if a pixel has pixels of different colors around itself. Given the finite resolution of the basins $\varepsilon$, these boundaries may appear slightly different even for Wada basins.

\item The slim boundaries $\partial B_i$ obtained in the previous step are fattened by fat pixels of fattening parameter $r$, becoming the fat boundaries $\boldsymbol{\partial B_i}$. We can start with $r=1$, and if the result of the Wada test in step (e) is negative, we can start over the step (d) with higher values of $r$ until we reach a stopping condition $r=r_{max}$.

\item We check if all the slim boundaries fit into all the fat boundaries $\partial B_i \subset \boldsymbol{\partial B_j}$ $\forall i,j=1,\ldots,N_a$. If this is the case, we say that the basins have the Wada property. Of course, this verification is reliable up to a resolution determined by the fattening parameter $r$. In case that the system is not Wada, the algorithm provides a list of non-Wada points of the basin.
\end{enumerate}

\subsection{\label{sec:Analysis}Analysis of the merging method}
The whole method described above relies on a single parameter: the fattening parameter $r$. This parameter determines the confidence that we have in the result of the algorithm, since we will be able to say that the basin is Wada up to the resolution defined by the fat pixels that we use. Therefore, it is natural to analyze the behavior of the method for different values of $r$ in dynamical systems with different features. This is exactly our aim in this section.


In order to examine the behavior of the procedure with respect to the fattening parameter $r$, we can apply it to different Wada boundaries. We can characterize fractal boundaries by their fractal dimension and by the number of basins that they separate at the same time. Here, we examine two dynamical systems with Wada boundaries where we can easily vary these two quantities. Namely, the two paradigmatic dynamical systems under study are the Hénon-Heiles Hamiltonian \cite{aguirre_wada_2001} and the Newton method to find complex roots \cite{epureanu_fractal_1998}.

\begin{figure*}
\begin{center}
\includegraphics[width=0.9\textwidth]{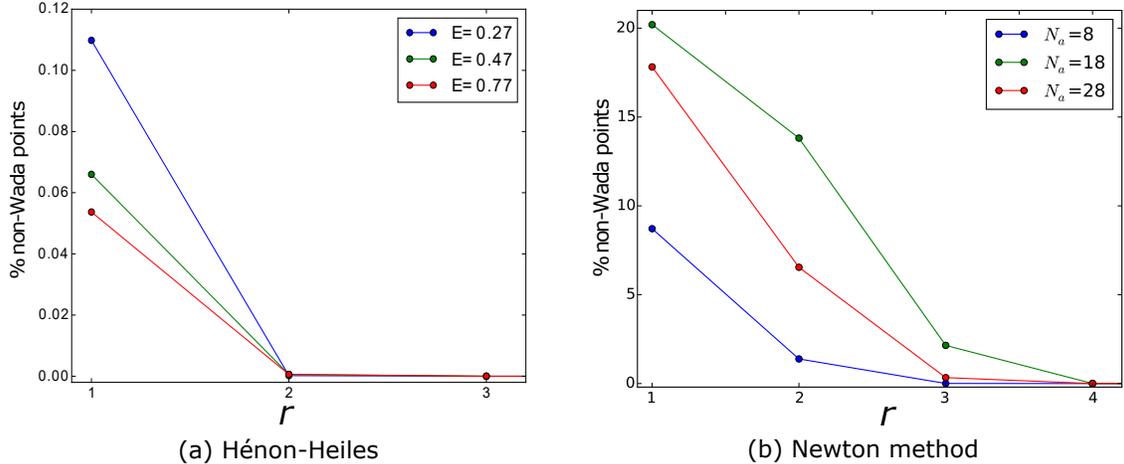}
\end{center}
\caption{\label{fig:merge_test} \textbf{The role of the fattening parameter $r$.} (a) The number of non-Wada points decreases as the fattening parameter $r$ increases for all the values of the energy studied in the H\'enon-Heiles Hamiltonian. Only three values of the energy are plotted for clarity, but we have checked that there is no relation between the percentage of non-Wada points and the value of the energy $E$, i.e., there is no relation between the percentage of non-Wada points and the fractal dimension of the boundary. (b) The merging method converges for values of the fattening parameter $r\leq4$ in the Newton method to find complex roots. There is no direct relation between the number of attractors and the percentage of non-Wada points. For both systems and all the parameters tested, the merging algorithm correctly classifies the basins as Wada basins for every $r>4$.}
\end{figure*}

The Hénon-Heiles Hamiltonian is defined by $H=\frac{1}{2}(\dot{x}^2+\dot{y}^2)+\frac{1}{2}(x^2+y^2)+x^2y-\frac{1}{3}y^3$. For values of the energy above the critical one, the escape basins of this open Hamiltonian system show the Wada property \cite{aguirre_wada_2001}. The fractal dimension of the boundaries diminishes as the energy increases, but the Wada property is preserved, as reported in \cite{blesa_escape_2012,zotos_overview_2017}. We have used the merging method described in the previous section with different values of the energy $E$ and of fattening parameter $r$. The results are plotted in Fig.~\ref{fig:merge_test}-(a). We have plotted only three different values of the energy for clarity, but we can observe that the algorithm correctly determines that the basins are Wada for $r\geq4$ at every tested value of the energy. It can also be noticed that there is no relation between the number of non-Wada points detected by the algorithm and the fractal dimension of the boundaries. Furthermore, we have carried out similar computations for increasing resolutions with analogous results. Thus, from these numerical experiments, we conclude that there is no relation between the value of the fattening parameter $r$ needed to correctly predict the Wada property and the fractal dimension of the boundaries.

The second system where we have tested the merging method described before is the Newton method to find complex roots. This method can be described by the discrete complex variable map $z_{n+1}=z_n-(z^R-1)/(R \cdot z^{R-1})$, where the parameter $R$ determines the number of roots and therefore the number of attractors $N_a=R$. It has been reported that the basins produced by this complex variable map show the disconnected Wada property (the basins are disconnected and also Wada) no matter the number of attractors determined by $R$ \cite{epureanu_fractal_1998, frame_newtons_2007, ziaukas_fractal_2017}. Thus, we ran the merging algorithm for an increasing number of roots $R$, and consequently of attractors $N_a$. As shown in Fig.~\ref{fig:merge_test}-(b), the merging method correctly classifies the basins as Wada for all $r>4$ even for a large number of attractors $N_a$, which seldom appears in typical dynamical systems. Moreover, we have found no trivial relation between the number of attractors $N_a$ and the percentage of non-Wada points. Again, we have performed the computations at different resolutions (up to $5000 \times 5000$) with consistent results. Hence, we conclude that the value of the fattening parameter $r$ needed for a correct classification of the basins does not directly depend on the fractal dimension of the boundaries nor on the number of attractors. This also proves that the method works correctly for disconnected Wada basins.

Finally, we would like to mention another adjustment that could be added to the merging algorithm in case of need. Just as described before, the merging algorithm is an all or nothing test. If there is a single pixel of a slim boundary that does not fit into a fat boundary, then the basin is labeled as non-Wada. However, it is clear that this can be too restrictive in some cases. For instance, if the basin is obtained by experimental procedures, it is very likely to have some wrong pixels. In these cases, we could complement the merging algorithm with the measure of the fractal dimension of the non-Wada boundary, using a box-counting algorithm on the resulting image of the non-Wada points, for example. If the fractal dimension of these non-Wada boundary points is close to zero, then we can admit that the basins have the Wada property, despite the misbehaved pixels. In any case, the merging method is able to determine whether a basin is Wada or not up to a given resolution using minimum requirements.

\section{\label{sec:Conclusions}Conclusions}

In the study of the asymptotic behavior of nonlinear dynamical systems, Wada basins appear frequently. Initial conditions lying in the boundary of Wada basins can suffer an arbitrarily small perturbation leading the trajectory to any of the possible attractors of the system. This supposes a special kind of unpredictability different not only from basins with smooth boundaries, but also from other fractal basins \cite{aguirre_fractal_2009, daza_basin_2016}.

In this paper, we have seen how the action of merging the basins reveals a new aspect of Wada basins. Actually, Wada boundaries are those that remain unaltered under the action of merging the basins. This perspective provides a new way to test Wada basins, that is faster than previous methods by orders of magnitude, and also much easier to use. Given a basin with three attractors with $1000\times1000$ initial conditions, it takes around two seconds to test the Wada property in a personal computer. Furthermore, no knowledge of the underlying dynamical system is needed. This means that this method is especially suitable for cases in which the exact equations and parameters governing the phenomenon are unknown. Besides, this method can be easily automated. Previous methods \cite{nusse_fractal_2000,daza_testing_2015} required a detailed knowledge of the dynamical system and important computational efforts. The only possible black spot of the merging method is that it tests the Wada property up to a given resolution (this is also true for the Wada test proposed in \cite{daza_testing_2015}). Nevertheless, the merging method is the best option to check the Wada property with minimum requirements. This is why we believe that the merging method will become a fundamental tool in the study of the Wada property with applications to many scientific and engineering contexts.

\section*{Acknowledgments}
We acknowledge some enlightening discussions with Jim Yorke. This work has been supported by the Spanish State Research Agency (AEI) and the European Regional Development Fund (FEDER) under Project No. FIS2016-76883-P. M.A.F.S. acknowledges the jointly sponsored financial support by the Fulbright Program and the Spanish Ministry of Education (Program No. FMECD-ST-2016).

\section*{Author contributions statement}
A.D., A.W., and M.A.F.S. devised the research. A.D. and A.W. performed the numerical simulations. A.D., A.W., and M.A.F.S. analyzed the results and wrote the paper.

\section*{Additional information}
\textbf{Competing interests}: The authors declare no competing interests.

\end{document}